\documentclass[journal,10pt]{IEEEtran}

\usepackage{cite}

\usepackage{amssymb}
\ifCLASSINFOpdf
   \usepackage[pdftex]{graphicx}
\else
   \usepackage[dvips]{graphicx}
\fi

\usepackage[cmex10]{amsmath}
\usepackage{mdwmath}

\usepackage{ntheorem}
{
\theoremheaderfont{\itshape}
\theorembodyfont{\upshape}
\theoremseparator{:}

\newtheorem{lemma}{Lemma}

\newtheorem*{proof}{Proof}
}

\usepackage{array}

\usepackage{multirow}  
\usepackage{chemarrow}

\ifCLASSOPTIONcompsoc
\usepackage[caption=false,font=normalsize,labelfont=sf,textfont=sf]{subfig}
\else
\usepackage[caption=false,font=footnotesize]{subfig}
\fi

\usepackage{stfloats}
\usepackage{extarrows}
\allowdisplaybreaks[4]

\usepackage{color}

\begin{document}


\title{Low-Interference {\it N}-Continuous OFDM via Optimized Time-Domain Smoothing}


\author{Peng~Wei, Yue~Xiao, Lilin~Dan, and Wei~Xiang
%
%
}
\maketitle

\begin{abstract}
A novel basis signal optimization method is proposed for reducing the interference in the \emph{N}-continuous orthogonal frequency division multiplexing (NC-OFDM) system.
Compared to conventional NC-OFDM, the proposed scheme is capable of improving the transmission performance while maintaining an identical sidelobe suppression performance imposed by the linear combination of two groups of basis signals.
Our performance results demonstrate that with a low-complexity overhead, the proposed scheme is capable of striking a better trade-off among the bit error rate (BER), complexity, and the sidelobe suppression performance compared to its conventional counterparts.
\end{abstract}


\IEEEpeerreviewmaketitle

\section{Introduction}

Since the proposal of \emph{N}-continuous orthogonal frequency division multiplexing (NC-OFDM) by Beek \emph{et al.} \cite{Ref1,Ref2}, a class of frequency-domain precoders has been developed to suppress out-of-band power emission by enforcing continuity of the transmitted signal up to the \emph{N}th-order derivative \cite{Ref1,Ref2,Ref3,Ref4,Ref5,Ref6,Ref7,Ref16,Ref17}.
Although the \emph{N}-continuity criterion has been widely employed for  sidelobe suppression, precoding in the frequency domain does not inherently guarantee the direct elimination of self-interference over multipath fading channels.

More specifically, the precoders in \cite{Ref1,Ref2} introduce direct self-interference that deteriorates the signal-to-noise ratio (SNR).
To alleviate this problem, selected mapping (SLM) \cite{Ref3}, cancellation tones \cite{Ref4}, grinder matrix \cite{Ref5}, error-vector magnitude (EVM)-constrained precoder \cite{Ref6}, and low-rank constraint matrix \cite{Ref7} have been employed, at the cost of computational complexity for interference-free reception.
Alternatively, the low-complexity orthogonal precoding architecture in \cite{Ref16,Ref17} requires an extra matrix multiplication at the receiver to reverse the orthogonal projection imposed at the transmitter.
A different approach in \cite{Ref18} avoids explicit signal recovery by exploiting a fast-decaying shaping function.
However, the resulting channel-dependent precoding matrix becomes sensitive to time-varying channels and demands frequent channel-state-information (CSI) feedback at high rates.

By contrast, the family of self-interference reduction precoders \cite{Ref8,Ref9} has been explicitly designed to suppress self-interference by confining the precoder-induced distortion to the guard interval.
As a result, these schemes outperform their counterparts in terms of sidelobe suppression, complexity, and bit error ratio (BER) performance.
Nevertheless, the interference added in the guard interval increases the sensitivity to channel-induced inter-symbol interference (ISI) and inter-carrier interference (ICI) \cite{Ref9}.
Moreover, the precoder in \cite{Ref8} requires knowledge of the previous symbol, which requires extra memory.

In recent years, studies of time-domain precoding have found reduced-complexity applications in \emph{N}-continuous OFDM \cite{Ref10,Ref11}.
More specifically, a linear-combination-based \emph{N}-continuous transmitter design that relies on basis signals has been proposed in \cite{Ref10,Ref11}.

Against this background, the novel contribution of this paper is a time-domain \emph{N}-continuous OFDM scheme aided by two linear-combination representations, which improves the continuity of the basis signals to adaptively control the self-interference without degrading the sidelobe suppression performance compared to conventional NC-OFDM.

\section{Time-Domain NC-OFDM}

In this section, we give a brief introduction of time-domain NC-OFDM \cite{Ref10,Ref11}, which with the given basis signals has an equivalent signal format as traditional NC-OFDM.
In a baseband OFDM system, the input bit stream of the \emph{i}th OFDM symbol is first modulated onto an uncorrelated complex-valued data vector $\mathbf{d}_i={[ d_{i,r_1}, d_{i,r_2},\ldots,d_{i,r_K}]}^{\rm T}$ drawn from a constellation such as phase-shift keying (PSK) or quadrature amplitude modulation (QAM). The complex-valued data vector is then mapped onto \emph{K} subcarriers with the index set $\mathcal{R}=\left\{r_1,r_2,\ldots,r_{K}\right\}$.
An OFDM signal is formed by summing all the \emph{K}-modulated orthogonal subcarriers with equal frequency spacing $\Delta f=1/T_{\rm s}$, where $T_{\rm s}$ is the OFDM symbol duration.
The oversampled signal, with a sampling interval of $T_{\rm samp}=T_{\rm s}/L_{\rm s}$, is then transformed into the discrete-time OFDM signal via an inverse discrete Fourier transform (IDFT), given by
\begin{equation}
\label{Eqn1}
x_i(l)=\frac{1}{L_{\rm s}}\sum\limits^{K}_{k=1}{d_{i,r_k}e^{j2\pi \frac{r_k}{L_{\rm s}}l}},
\end{equation}
where $l \in \mathcal{L} \triangleq \left\{-L_{\rm cp},-L_{\rm cp}+1,\ldots,L_{\rm s}-1\right\}$ and $L_{\rm cp}$ denotes the length of the cyclic prefix (CP).

According to \cite{Ref10,Ref11} for time-domain \emph{N}-continuous OFDM, a smooth signal $\tilde{w}_i(l)$ is directly added to the CP-appended OFDM symbol $x_i(l)$ for $l \in \mathcal{L}$ to construct the \emph{N}-continuous signal $\bar{x}_i(l)$ given by
\begin{equation} \label{Eqn2}
\bar{x}_i(l)=x_i(l)+\tilde{w}_i(l).
\end{equation}

Aiming at eliminate discontinuities at the adjacent point indexed by $-L_{\rm cp}$ between two consecutive OFDM symbols as shown in Fig. \ref{fig:fig1}, $\tilde{w}_i(l)$ should follow
\begin{equation}\label{Eqn3}
\left.\tilde{w}^{(n)}_i(l)\right|_{l=-L_{\rm cp}} \!=\! \left.\bar{x}^{(n)}_{i-1}(l)\right|_{l=L_{\rm s}} \!-\! \left.x^{(n)}_i(l)\right|_{l=-L_{\rm cp}}\!,
\end{equation}
where $n\in\mathcal{U}_N \triangleq \{0,1,\ldots,N\}$ and $x^{(n)}_i(l)$ represents the $n$th derivative of $x_i(l)$. Furthermore, $\tilde{w}_i(l)$ is designed by linearly combining a group of basis signals $f^{(n)}(l)$ as follows
\begin{equation} \label{Eqn4}
\tilde{w}_i(l)=\sum\limits^{N}_{n=0} b_{i,n} f^{(n)}(l),
\end{equation}
where $l \in \mathcal{L}$ and $f^{(n)}(l)$ has the following design as
\begin{equation}  \label{Eqn5}
f^{(n)}(l)=   \left( \frac{j 2\pi}{L_{\rm s}}\right)^{{n}}\sum\limits_{r_k\in\mathcal{R}}{r^{n}_k e^{j2\pi\frac{r_k}{L_{\rm s}}(l+L_{\rm cp})}}.
\end{equation}

Since the support length of $f^{(n)}(l)$ is the same as that of $x_i(l)$, the interference caused by $\tilde{w}_i(l)$, existed in the whole OFDM symbol, results in severe performance degradation in BER. To solve this problem, an interference reduction design will be proposed via optimizing the basis signals.

\section{Optimized Time-Domain Smoothing at the OFDM Transmitter}

\begin{figure}[htpb]
\centering
\includegraphics[width=3.5in]{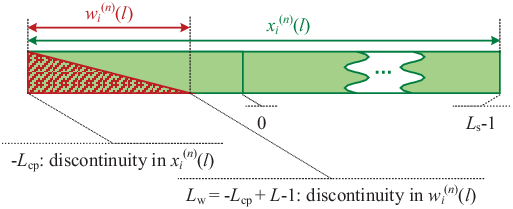}
\caption{Illustration of derivative discontinuities at two points in a baseband-equivalent OFDM symbol for low-interference \emph{N}-continuous signaling.}
\label{fig:fig1}
\end{figure}

As illustrated in Fig. \ref{fig:fig1}, by shortening the length of the smooth signal from $L_{\rm s}+L_{\rm cp}$ to $L$ $(0<L<L_{\rm s}+L_{\rm cp})$, the optimized smooth signal $w_i(l)$ is overlapped with $x_i(l)$ according to \eqref{Eqn2} to construct the \emph{N}-continuous signal $\bar{x}_i(l)$.

Different from the traditional time-domain NC-OFDM system with discontinuities of $x^{(n)}_{i}(l)$ for $n\in\mathcal{U}_N$ at just one point, since the length of $w_i(l)$ is smaller than that of $x_i(l)$, two disjoint points are required to be smoothed in the proposed scheme.
As depicted in Fig. \ref{fig:fig1}, the first one indexed by $-L_{\rm cp}$ and the second one indexed by $L_{\rm w} = -L_{\rm cp}+ L-1$ are introduced, which correspond to the beginning and end of $w_i(l)$, respectively. Thus, to meet the \emph{N}-continuity criterion, the discontinuities at the two disjoint points are required to be smoothed at the same time.

At index $-L_{\rm cp}$, since the previous symbol $x_{i-1}(l)$ is not affected by $w_{i-1}(l)$, for $n\in\mathcal{U}_N$, $w_i(l)$ should satisfy
\begin{equation} \label{Eqn6}
\left.w^{(n)}_i(l)\right|_{l=-L_{\rm cp}} \!=\! \left.x^{(n)}_{i-1}(l)\right|_{l=L_{\rm s}} \!-\! \left.x^{(n)}_i(l)\right|_{l=-L_{\rm cp}}.
\end{equation}
Simultaneously, at index $L_{\rm w}$, for the discontinuities caused by $w^{(n)}_i(l)$ for $n\in\mathcal{U}_N$, $w_i(l)$ should satisfy
\begin{equation} \label{Eqn7}
\left.{w}^{(n)}_i(l)\right|_{l=L_{\rm w}} = 0.
\end{equation}

Here we focus on designing $w_i(l)$ satisfying \eqref{Eqn6} and \eqref{Eqn7}. Inspired by \eqref{Eqn4}, to eliminate the $N+1$ discontinuities at time $-L_{\rm cp}$ or $L_{\rm w}$, each disjoint point requires at least $N+1$ basis signals. Thus, in our proposed scheme, $w_i(l)$ including two linear combinations of two groups of basis signals is formulated as
\begin{equation} \label{Eqn8}
w_i(l)=\! \left\{\begin{array}{cc}
 \!\!\!\! \sum\limits^{N}_{n=0}{ b_{i,n} f_n(l)} \!+\! \! \sum\limits^{N}_{n=0}{ \tilde{b}_{i,n} \tilde{f}_n(l)}, &\! l\in\mathcal{W}, \\
0, & \! l\in \mathcal{L} \setminus \mathcal{W},
\end{array}\right.
\end{equation}
where the two linear combinations correspond to \eqref{Eqn6} and \eqref{Eqn7}, respectively.
$\mathcal{L} \setminus \mathcal{W} \triangleq \{l \mid l \in \mathcal{L} \;\text{and}\; l \notin \mathcal{W}\}$.
$\mathcal{W}\triangleq \{ -L_{\rm cp},-L_{\rm cp}+1, \ldots, L_{\rm w}\}$ indicates the range of $w_i(l)$, which implies that the self-interference caused by $w_i(l)$ can be adaptively adjusted by $L$. 
Furthermore, the basis signals ${f}_n(l)$ and $\tilde{f}_n(l)$ are, respectively, designed as
\begin{equation} \label{Eqn9}
  f_n(l)=g_n(l) s_1(l) ,
\end{equation}
\begin{equation} \label{Eqn10}
  \tilde{f}_n(l)=\tilde{g}_n(l) s_2(l) ,
\end{equation}
which belong to basis sets ${\mathcal{Q}}$ and $\tilde{\mathcal{Q}}$, respectively, as follows
\begin{align}  \label{Eqn11}
  {\mathcal{Q}} \triangleq \left\{
   {\mathbf{q}}_{{n}}\left|{\mathbf{q}}_{{n}}
  = \left[ {f}_{{n}}(-L_{\rm cp}), \ldots, {f}_{{n}}(L_{\rm w})\right]^{\rm T}\right., {n}\in \mathcal{U}_{N}  \right\},
  \end{align}
\begin{align}  \label{Eqn12}
  \tilde{\mathcal{Q}} \triangleq  \left\{
  \tilde{\mathbf{q}}_{{n}}\left|\tilde{\mathbf{q}}_{{n}}
  = \left[\tilde{f}_{{n}}(-L_{\rm cp}),\ldots, \tilde{f}_{{n}}(L_{\rm w}) \right]^{\rm T}\right., {n}\in \mathcal{U}_{N} \right\},
  \end{align}
where according to \eqref{Eqn5}, $g_n(l)$ and $\tilde{g}_{{n}}(l)$ are given by
\begin{equation}  \label{Eqn13}
g_{{n}}(l)= f^{(n)}(l),
\end{equation}
\begin{equation}  \label{Eqn14}
\tilde{g}_{{n}}(l)= f^{(n)}(l- L), 
\end{equation}
$s_1(l)$ and $s_2(l)$ are the truncation windows.

It is noted that when a zero-edge window function $s(l)$ whose first $N_1$ derivatives are continuous is adopted as the truncation window, the continuities of ${f}_n(l)$ and $\tilde{f}_n(l)$ and their first $N_1$ derivatives will be enhanced at the truncated points corresponding to the end and beginning of $w_i(l)$. This further smoothes the discontinuities of point $w^{(n)}_i(L_{\rm w})$ for $n=0,1,\ldots,N_1$. In this case, the number of discontinuities required to be removed at time $L_{\rm w}$ is reduced from $N+1$ to $N-N_1$. 
Thus, to remove the remaining $N-N_1$ discontinuities, the design of $w_i(l)$ in \eqref{Eqn8} is varied to
\begin{equation} \label{Eqn16}
w_i(l)\!=\!\! \left\{\begin{array}{cc}
 \!\!\!\! \sum\limits^{N}_{n=0}\!{ b_{i,n} f_n(l)} \! + \!\!\! \sum\limits^{\tilde{N}}_{n=0} { \tilde{b}_{i,n} \tilde{f}_n(l)}, & \!\! l \!\in \! \mathcal{W}, \\
0, &\!\! l \! \in \! \mathcal{L} \setminus \mathcal{W}.
\end{array}\right.
\end{equation}
where $\tilde{N} = N-N_1-1$.
Using the above $s(l)$ for $l = 0,1,\ldots, 2L-2$ yields the truncation windows as
\begin{equation}  \label{Eqn17}
  s_1(l)=s(l+L_{\rm cp}+L_{\rm w}) u(l+L_{\rm cp}), \; l \in \mathcal{W},
\end{equation}
\begin{equation}  \label{Eqn18}
  s_2(l)=s(l+L_{\rm cp}) u(-l+L_{\rm w}), \; l\in \mathcal{W},
\end{equation}
$s_1(l)=0$ and $s_2(l)=0$ for $l \notin \mathcal{W}$, where $u(l)$ is the unit step function. Actually, $s_1(l)$ and $s_2(l)$ correspond to the right and left halves of $s(l)$, respectively.

Then, based on the carefully designed $f_n(l)$ and $\tilde{f}_{{n}}(l)$, coefficients $b_{i,n}$ and $\tilde{b}_{i,n}$ are calculated to generate $w_i(l)$ as follows.
Firstly, Eq. \eqref{Eqn16} is rewritten in matrix form as
\begin{equation}
\mathbf{w}_i= \mathbf{Q}_f \mathbf{b}_i +  \mathbf{Q}_{\tilde{f}} \tilde{\mathbf{b}}_i,
  \label{Eqn:Eqn19}
\end{equation}
where according to \eqref{Eqn11} and \eqref{Eqn12},  $\mathbf{Q}_f$ and $\mathbf{Q}_{\tilde{f}}$ are, respectively, expressed as
\begin{equation} \label{Eqn36}
	\mathbf{Q}_f\!=\!\begin{bmatrix}
		f_0(-\!L_{\rm cp}) & f_1(-\!L_{\rm cp}) & \cdots & f_N(-\!L_{\rm cp}) \\
		f_0(-\!L_{\rm cp}\!+\!1) & f_1(-\!L_{\rm cp}\!+\!1) & \cdots & f_N(-\!L_{\rm cp}\!+\!1) \\
		\vdots & \vdots & {} & \vdots \\
		f_0(L_{\rm w}) & f_1(L_{\rm w}) & \cdots & f_N(L_{\rm w}) \\
	\end{bmatrix}
\end{equation}
and
\begin{equation} \label{Eqn37}
	\mathbf{Q}_{\tilde{f}}\!=\!\! \begin{bmatrix}
		\tilde{f}_0(-\!L_{\rm cp}) \!\!\!& \tilde{f}_1(-\!L_{\rm cp}) \!\!\!& \cdots \!\!\!& \tilde{f}_{\tilde{N}}(-\!L_{\rm cp}) \\
		\tilde{f}_0(-\!L_{\rm cp}\!+\!1) \!\!\!& \tilde{f}_1(-\!L_{\rm cp}\!+\!1) \!\!\!& \cdots \!\!\!& \tilde{f}_{\tilde{N}}(-\!L_{\rm cp}\!+\!1) \\
		\vdots \!\!\!& \vdots \!\!\!& {} \!\!\!& \vdots \\
		\tilde{f}_0(L_{\rm w}) \!\!\!& \tilde{f}_1(L_{\rm w}) \!\!\!& \cdots \!\!\!& \tilde{f}_{\tilde{N}}(L_{\rm w}) \\
	\end{bmatrix}\!\!.
\end{equation}

Secondly, to satisfy both conditions \eqref{Eqn6} and \eqref{Eqn7}, substituting \eqref{Eqn:Eqn19} into \eqref{Eqn6} and \eqref{Eqn7} yields $2N-N_1+1$ equations, expressed in the following matrix form
\begin{equation}  \label{Eqn19}
\begin{bmatrix} \mathbf{P}_f & \mathbf{P}_{\tilde{f}}\\ \mathbf{P}_{f_{\rm w}} &  \mathbf{P}_{\tilde{f}_{\rm w}} \end{bmatrix}
\begin{bmatrix} \mathbf{b}_i  \\ \tilde{\mathbf{b}}_i \end{bmatrix}
=\begin{bmatrix} \Delta\mathbf{x}_i  \\ \mathbf{0}_{(N+1)\times 1} \end{bmatrix},
\end{equation}
where $\mathbf{b}_i=\left[  {b}_{i,0}, b_{i,1} , \ldots, b_{i,N} \right]^{\rm T}$,
and $\tilde{\mathbf{b}}_i   =  [  \tilde{b}_{i,0}, \tilde{b}_{i,1} , \ldots, \tilde{b}_{i,\tilde{N}} ]^{\rm T}$, and matrices $\mathbf{P}_{{f}}$, $\mathbf{P}_{\tilde{f}}$, $\mathbf{P}_{{f}_{\rm w}}$, and $\mathbf{P}_{\tilde{f}_{\rm w}}$ are expressed as
$$\mathbf{P}_f=\!\!\begin{bmatrix}
{f}_0(-L_{\rm cp}) & {f}_1(-L_{\rm cp}) &  \cdots &  {f}_N(-L_{\rm cp}) \\
{f}^{(1)}_0(-L_{\rm cp}) & {f}^{(1)}_1(-L_{\rm cp}) &  \cdots &  {f}^{(1)}_N(-L_{\rm cp}) \\
\vdots  & \vdots  & {} &\vdots\\
{f}^{(N)}_0(-L_{\rm cp}) & {f}^{(N)}_1(-L_{\rm cp}) &  \cdots &  {f}^{(N)}_N(-L_{\rm cp}) \\
\end{bmatrix}\!\!,$$
$$\mathbf{P}_{\tilde{f}}=\!\! \begin{bmatrix}
\tilde{f}_0(-L_{\rm cp}) & \tilde{f}_1(-L_{\rm cp}) &  \cdots &  \tilde{f}_{\tilde{N}}(-L_{\rm cp}) \\
\tilde{f}^{(1)}_0(-L_{\rm cp}) & \tilde{f}^{(1)}_1(-L_{\rm cp}) &  \cdots &  \tilde{f}^{(1)}_{\tilde{N}}(-L_{\rm cp}) \\
\vdots  & \vdots  & {} &\vdots\\
\tilde{f}^{(N)}_0(-L_{\rm cp}) & \tilde{f}^{(N)}_1(-L_{\rm cp}) &  \cdots &  \tilde{f}^{(N)}_{\tilde{N}}(-L_{\rm cp}) \\
\end{bmatrix}\!\!,$$
$$ \mathbf{P}_{f_{\rm w}}=\!\! \begin{bmatrix}
{f}_0(L_{\rm w}) & {f}_1(L_{\rm w}) &  \cdots &  {f}_N(L_{\rm w}) \\
{f}^{(1)}_0(L_{\rm w}) & {f}^{(1)}_1(L_{\rm w}) &  \cdots &  {f}^{(1)}_N(L_{\rm w}) \\
\vdots  & \vdots  & {} &\vdots\\
{f}^{(N)}_0(L_{\rm w}) & {f}^{(N)}_1(L_{\rm w}) &  \cdots &  {f}^{(N)}_N(L_{\rm w}) \\
\end{bmatrix}\!\!, $$
$$ \mathbf{P}_{\tilde{f}_{\rm w}} \!\!=\!\! \begin{bmatrix}
\tilde{f}_0(L_{\rm w}) \!\!\!& \tilde{f}_1(L_{\rm w}) \!\!\!&  \cdots \!\!\!&  \tilde{f}_{\tilde{N}}(L_{\rm w}) \\
\tilde{f}^{(1)}_0(L_{\rm w}) \!\!\!& \tilde{f}^{(1)}_1(L_{\rm w}) \!\!\!&  \cdots \!\!\!&  \tilde{f}^{(1)}_{\tilde{N}}(L_{\rm w}) \\
\vdots  \!\!\!& \vdots  \!\!\!& {} \!\!\!&\vdots\\
\tilde{f}^{(N)}_0(L_{\rm w}) \!\!\!& \tilde{f}^{(N)}_1(L_{\rm w}) \!\!\!&  \cdots \!\!\!&  \tilde{f}^{(N)}_{\tilde{N}}(L_{\rm w}) \\
\end{bmatrix}\!\!.$$ %
$\Delta\mathbf{x}_i$ is an $(N+1)\times 1$ vector of the differences between $\mathbf{x}_i$ and $\mathbf{x}_{i-1}$, as well as their first $N$ derivatives at index $-L_{\rm cp}$, which can be calculated by
\begin{equation}  \label{Eqn20}
  \Delta\mathbf{x}_i= {\mathbf{P}}_1\mathbf{d}_{i-1}-\mathbf{P}_2\mathbf{d}_i,
\end{equation}
where 
\begin{equation}
{\mathbf{P}}_1 \!=\! \frac{1}{L_{\rm s}} \! \begin{bmatrix}  \label{Eqn21}
1  & 1  & \cdots &  1 \\
\dfrac{j2\pi r_1}{L_{\rm s}}  &\dfrac{j2\pi r_2}{L_{\rm s}}  & \cdots &\dfrac{j2\pi r_K}{L_{\rm s}} \\
\vdots  & {} &{} &\vdots\\
\left(\!\dfrac{j2\pi r_1}{L_{\rm s}}\!\right)^{\!\! N}  & \left(\!\dfrac{j2\pi r_2}{L_{\rm s}}\!\right)^{\!\! N}  & \cdots &  \left(\!\dfrac{j2\pi r_K}{L_{\rm s}}\!\right)^{\!\! N} \\
\end{bmatrix}\!\!,
\end{equation}
\begin{equation} \label{Eqn22}
\mathbf{P}_{2}={\mathbf{P}}_1\mathbf{U},
\end{equation}
$\mathbf{U}\triangleq {\rm diag}\{e^{j\varphi r_1},e^{j\varphi r_2},\ldots,e^{j\varphi r_K} \}$ is a $K \times K$ diagonal matrix, and $\varphi=-2\pi L_{\rm cp}/L_{\rm s}$.
Lastly, when the rank of the coefficient matrix $\begin{bmatrix} \mathbf{P}_f &\!\! \mathbf{P}_{\tilde f} \\  \mathbf{P}_{f_{\rm w}} &\!\! \mathbf{P}_{{\tilde f}_{\rm w}} \end{bmatrix}$ is equal to that of the augmented matrix $\begin{bmatrix} \mathbf{P}_f &\!\! \mathbf{P}_{\tilde f} &\!\! \Delta\mathbf{x}_i \\  \mathbf{P}_{f_{\rm w}} &\!\! \mathbf{P}_{{\tilde f}_{\rm w}} &\!\! \mathbf{0}_{(N+1)\times 1}  \end{bmatrix}$  in \eqref{Eqn19}, if the coefficient matrix is of full rank, the coefficient vectors can be derived as
\begin{eqnarray}
\begin{bmatrix} \mathbf{b}_i \\ \tilde{\mathbf{b}}_i \end{bmatrix} & \!\!=\!\! & \begin{bmatrix} \mathbf{P}_f & \mathbf{P}_{\tilde f} \\  \mathbf{P}_{f_{\rm w}} & \mathbf{P}_{{\tilde f}_{\rm w}} \end{bmatrix}^{-1}  \begin{bmatrix} {\mathbf{P}}_1\mathbf{d}_{i-1}-\mathbf{P}_2\mathbf{d}_i \\  \mathbf{0}_{(N+1)\times 1}  \end{bmatrix}  \label{Eqn23} \\
{} & \!\!=\!\! & \mathbf{P}  \left( {\mathbf{P}}_1\mathbf{d}_{i-1}-\mathbf{P}_2\mathbf{d}_i \right),
\label{Eqn24}
\end{eqnarray}
where $\mathbf{P}$ is composed of the first $N+1$ columns of $\begin{bmatrix} \mathbf{P}_f &\!\!\! \mathbf{P}_{\tilde f} \\  \mathbf{P}_{f_{\rm w}} &\!\!\! \mathbf{P}_{{\tilde f}_{\rm w}} \end{bmatrix}^{\!-1}$.
Additionally, under the condition of \eqref{Eqn23}, if the inverse of matrix $\mathbf{P}_f - \mathbf{P}_{\tilde{f}} \mathbf{P}^{-1}_{\tilde{f}_{\rm w}} \mathbf{P}_{f_{\rm w}}$ exists, Eq. \eqref{Eqn23} can also be expressed as
\begin{subequations}
\begin{align}
& {\mathbf{b}}_i=( \mathbf{P}_f - \mathbf{P}_{\tilde{f}} \mathbf{P}^{-1}_{\tilde{f}_{\rm w}} \mathbf{P}_{f_{\rm w}}  )^{-1} \Delta\mathbf{x}_i,   \label{Eqn25a} \\
& \tilde{\mathbf{b}}_i = -\mathbf{P}^{-1}_{\tilde{f}_{\rm w}} \mathbf{P}_{f_{\rm w}} ( \mathbf{P}_f - \mathbf{P}_{\tilde{f}} \mathbf{P}^{-1}_{\tilde{f}_{\rm w}} \mathbf{P}_{f_{\rm w}}  )^{-1} \Delta\mathbf{x}_i. \label{Eqn25b}
\end{align}
 \label{Eqn25}
\end{subequations}
Otherwise, if the inverse of matrix $ \mathbf{P}_{\tilde{f}_{\rm w}} - \mathbf{P}_{f_{\rm w}} \mathbf{P}^{-1}_f \mathbf{P}_{\tilde{f}}$ exists, Eq. \eqref{Eqn23} can be expressed as
\begin{subequations}
\begin{align}
& {\mathbf{b}}_i=\mathbf{P}^{-1}_f  ( \mathbf{I}_{N+1} \!+ \! \mathbf{P}_{\tilde{f}} ( \mathbf{P}_{\tilde{f}_{\rm w}} \!-\! \mathbf{P}_{f_{\rm w}} \mathbf{P}^{-1}_f \mathbf{P}_{\tilde{f}}  )^{\!-1}  \mathbf{P}_{f_{\rm w}} \mathbf{P}^{-1}_f )  \Delta\mathbf{x}_i,   \label{Eqn26a} \\
& \tilde{\mathbf{b}}_i = -( \mathbf{P}_{\tilde{f}_{\rm w}} - \mathbf{P}_{f_{\rm w}} \mathbf{P}^{-1}_f \mathbf{P}_{\tilde{f}}  )^{-1} \mathbf{P}_{f_{\rm w}} \mathbf{P}^{-1}_f  \Delta\mathbf{x}_i. \label{Eqn26b}
\end{align}
\label{Eqn26}
\end{subequations}
It is noted that when the rank of the coefficient matrix is smaller than $2N-N_1+1$ and equal to that of the augmented matrix, there will be infinite solutions in \eqref{Eqn19}. In this case, the inverse of the coefficient matrix in \eqref{Eqn23} is changed to the Moore-Penrose pseudo-inverse for a minimum Euclidean-norm solution.

Finally, it follows from \eqref{Eqn2}, \eqref{Eqn16}, and \eqref{Eqn24} that for $M$ consecutive OFDM symbols, i.e., $1\leq i  \leq M$, the \emph{N}-continuous symbol $\bar{\mathbf{x}}_i$ is expressed as
\begin{equation}
  \bar{\mathbf{x}}_i\!=\!\left\{\begin{matrix}
      \mathbf{x}_i \!+\!   \begin{bmatrix}  \mathbf{Q}_{{f}}  &\!\! \mathbf{Q}_{\tilde{f}}  \end{bmatrix}
       \mathbf{P}  \left( {\mathbf{P}}_1\mathbf{d}_{i-1} \!-\! \mathbf{P}_2\mathbf{d}_i \right) \!, & l \!\in\! \mathcal{W},\\ 
       \mathbf{x}_i, & l \!\in\! \mathcal{L} \!\setminus\! \mathcal{W}, 
\end{matrix}\right.
  \label{Eqn27}
\end{equation}
where $\mathbf{x}_i=[x_i(-L_{\rm cp}),x_i(-L_{\rm cp}+1),\ldots,x_i(L_{\rm s}-1)]^{\rm T}$ and $\mathbf{d}_{0}$ is initialized as a zero vector $\mathbf{d}_{0}=\mathbf{0}_{K\times 1}$.

\section{Interference Cancellation at the Receiver}

At the transmitter, when the length of the smooth signal is short, the interference to the useful information is small.
On the contrary, when the length of the smooth signal becomes longer, the interference will increase to degrade the system performance.
In this case, to eliminate the interference, it should be quantified to evaluate the effect to the transmission performance.
Thus, in this section, we first evaluate the SNR of the received signal to show the effect of the smooth signal to the useful information.
Then, to reduce the interference, an iterative signal recovery algorithm is proposed.

When the perfect synchronization is assumed and the CP can resist the effect of the multipath channel, after removing the CP, the discrete-time received signal $\mathbf{y}_i$ is given by
\begin{align}\label{Eqn29}
  \mathbf{y}_i 
  &= \bar{\mathbf{H}}_i \bar{\mathbf{x}}_i + \mathbf{z}_i \\
  &= \left[ \mathbf{0}_{L_{\rm s} \times (L_{\rm cp}-\tau)} \;\; \mathbf{H}_i \right] \bar{\mathbf{x}}_i + \mathbf{z}_i,
\end{align}
where $\mathbf{0}_{L_{\rm s} \times (L_{\rm cp}-\tau)}$ is an $L_{\rm s} \times (L_{\rm cp}-\tau)$ zero matrix, 
$\mathbf{y}_i=[y(0), y(1),\ldots,y(L_{\rm s}-1)]^{\rm T}$, the $L_{\rm s} \times (L_{\rm s}+\tau)$ channel matrix $\mathbf{H}_i$ is expressed as
\begin{equation}\label{Eqn30}
 \mathbf{H}_i=\begin{bmatrix}
      h_{i,\tau} & \cdots & h_{i,0} & 0 & \cdots & \cdots & 0  \\
      0 & h_{i,\tau} & \cdots & h_{i,0} & 0 & \cdots & 0 \\
       \vdots &   & & \ddots  &   & & \vdots \\
       0 & \cdots & \cdots &  0 & h_{i,\tau} & \cdots & h_{i,0} \\
    \end{bmatrix}\!\!,
\end{equation}
$\tau$ denotes the maximal channel delay in the discrete time and $\mathbf{z}_i$ is the additive Gaussian white noise (AWGN) with mean zero and variance $\sigma^2_{z}$.

According to \eqref{Eqn2} and \eqref{Eqn16}, Eq. \eqref{Eqn29} is further expressed as
\begin{eqnarray}
   \mathbf{y}_i
   &=& \bar{\mathbf{H}}_i \left( \mathbf{x}_i + \begin{bmatrix} \mathbf{w}_i \\ \mathbf{0}_{(L_{\rm s}+L_{\rm cp}-L)\times 1} \end{bmatrix}\right)+ \mathbf{z}_i   \label{Eqn31} \\
   &=& \frac{1}{L_{\rm s}} \mathbf{H}_{d_i} \mathbf{F}^{\rm H} \mathbf{d}_i + \mathbf{H}_{w_i} \mathbf{w}_i + \mathbf{z}_i,  \label{Eqn32}
\end{eqnarray}
where $\mathbf{F}$ is the discrete Fourier transform (DFT) matrix with the element of $\{\mathbf{F}\}_{k,(l+1)} \! = \! e^{-j2\pi \frac{r_k}{L_{\rm s}}l}$ for $k=1,2,\ldots,K$ and $l=0,1,\ldots,L_{\rm s}-1$, $\mathbf{H}_{d_i}$ is an $L_{\rm s} \times L_{\rm s}$ circulant matrix as
\begin{align}\label{Eqn33}
 &  \mathbf{H}_{d_i}
    = \left[ \mathbf{0}_{L_{\rm s} \times (L_{\rm cp}-\tau)} \;\; \mathbf{H}_i \right] \begin{bmatrix} \begin{bmatrix} \mathbf{0}_{L_{\rm cp} \times (L_{\rm s}-L_{\rm cp})} & \mathbf{I}_{L_{\rm cp}} \end{bmatrix} \\  \mathbf{I}_{L_{\rm s}} \end{bmatrix}   \nonumber \\
   &=\begin{bmatrix}
   h_{i,0} & 0 & \cdots  & 0 & h_{i,\tau} & \cdots & \cdots& \cdots & h_{i,1} \\
   h_{i,1} & h_{i,0} & 0  & \cdots &  0 & h_{i,\tau} & \cdots & \cdots & h_{i,2} \\
   \vdots & & & & &  \ddots & & & \vdots \\
   0 & \cdots  & \cdots & \cdots & \cdots & 0 & h_{i,\tau} & \cdots & h_{i,0}
   \end{bmatrix} \!.
\end{align}
The structure of the $L_{\rm s} \times L$ matrix $\mathbf{H}_{w_i}$ depends on the length $L$ of $\mathbf{w}_i$. 
If $L \le L_{\rm cp}-\tau$, then $\mathbf{H}_{w_i}$ is an all-zero matrix.
On the contrary, when $L > L_{\rm cp} - \tau$, it takes the form
\begin{align}\label{Eqn34}
   \mathbf{H}_{w_i} 
   &= \left[ \mathbf{0}_{L_{\rm s} \times (L_{\rm cp}-\tau)} \;\; \mathbf{H}_i \right]  \begin{bmatrix} \mathbf{I}_{L} \\ \mathbf{0}_{(L_{\rm s}+L_{\rm cp}-L) \times L} \end{bmatrix} \nonumber \\
   & =  \mathbf{H}_i   \begin{bmatrix} \begin{bmatrix} \mathbf{0}_{(L-L_{\rm cp}+\tau) \times (L_{\rm cp}-\tau)} & \mathbf{I}_{L-L_{\rm cp}+\tau} \end{bmatrix}  \\ \mathbf{0}_{(L_{\rm s}+L_{\rm cp}-L) \times L} \end{bmatrix}.  
\end{align}

\subsection{SNR Analysis}

The average SNR $\gamma$ at the receiver can be expressed as
\begin{align}\label{Eqn38}
   \gamma
   &=\frac{E\left\{ \left( \bar{\mathbf{H}}_i \mathbf{x}_i \right)^{\rm H} \bar{\mathbf{H}}_i \mathbf{x}_i  \right\}}{E\left\{ \!\! \left( \bar{\mathbf{H}}_i \! \begin{bmatrix} \mathbf{w}_i \\ \mathbf{0}_{(L_{\rm s}\!-\!L_{\rm w})\times 1} \end{bmatrix} \! \right)^{\!\rm H} \bar{\mathbf{H}}_i \! \begin{bmatrix} \mathbf{w}_i \\ \mathbf{0}_{(L_{\rm s}\!-\!L_{\rm w})\times 1} \end{bmatrix} \! \right\} \!+\! \sigma^2_z}.
\end{align}

We assume that the uncorrelated data $d_{i,r_k}$ follows $E\{\mathbf{d}_i \mathbf{d}^{\rm H}_i\}=\mathbf{I}_{K}$ and $E\{\mathbf{d}_i \mathbf{d}^{\rm H}_{i-1}\}=\mathbf{0}_{K \times K}$.
Substituting \eqref{Eqn22}, \eqref{Eqn24}, and \eqref{Eqn31}-\eqref{Eqn34} into \eqref{Eqn38} yields
\begin{align}\label{Eqn39}
   \gamma
   &=\frac{ \frac{1}{L^2_{\rm s}} {\rm Tr} \left\{ E\left\{ \mathbf{H}_{d_i} \mathbf{F}^{\rm H} \mathbf{d}_i \mathbf{d}^{\rm H}_i \mathbf{F} \mathbf{H}^{\rm H}_{d_i} \right\} \right\}}{ {\rm Tr} \left\{E\left\{ \mathbf{H}_{w_i} \mathbf{w}_i \mathbf{w}^{\rm H}_i \mathbf{H}^{\rm H}_{w_i} \right\}\right\} + \sigma^2_z}  \nonumber \\
      &=\frac{   \frac{1}{L_{\rm s}} \sum\limits^{K}_{k=1} E\left\{ |\lambda_{i,r_k}|^2 \right\}}
          { 2 {\rm Tr} \! \left\{\!E\!\left\{ \mathbf{H}^{\rm H}_{w_i} \mathbf{H}_{w_i} \right\} [\mathbf{Q}_f \; \mathbf{Q}_{\tilde{f}} ] \mathbf{P} \mathbf{P}_1 \mathbf{P}^{\rm H}_{1} \mathbf{P}^{\rm H} \! \begin{bmatrix} \mathbf{Q}^{\rm H}_f \\ \mathbf{Q}^{\rm H}_{\tilde{f}} \end{bmatrix}\! \right\} \!+\! \sigma^2_z}.
\end{align}
where $\lambda_{i,r_k}$ is the frequency-domain channel response.
The derivation of \eqref{Eqn39} is shown in Appendix \ref{app:app1}.

Fig. \ref{fig:fig2} compares the analytical and simulated SNRs over multipath fading channels, using the simulation parameters specified in Section V. 
The analytical curves closely match the simulation results. 
Moreover, for large $N$ and long $L$ (e.g., $N=4,6$ and $L=1024$), a moderate SNR loss is observed in the high ${\rm E_b/N_0}$ regime. 
By contrast, for small $N$ or short $L$, only a slight SNR loss occurs in the same region.

\begin{figure}[h]
\centering
\includegraphics[width=3.5in]{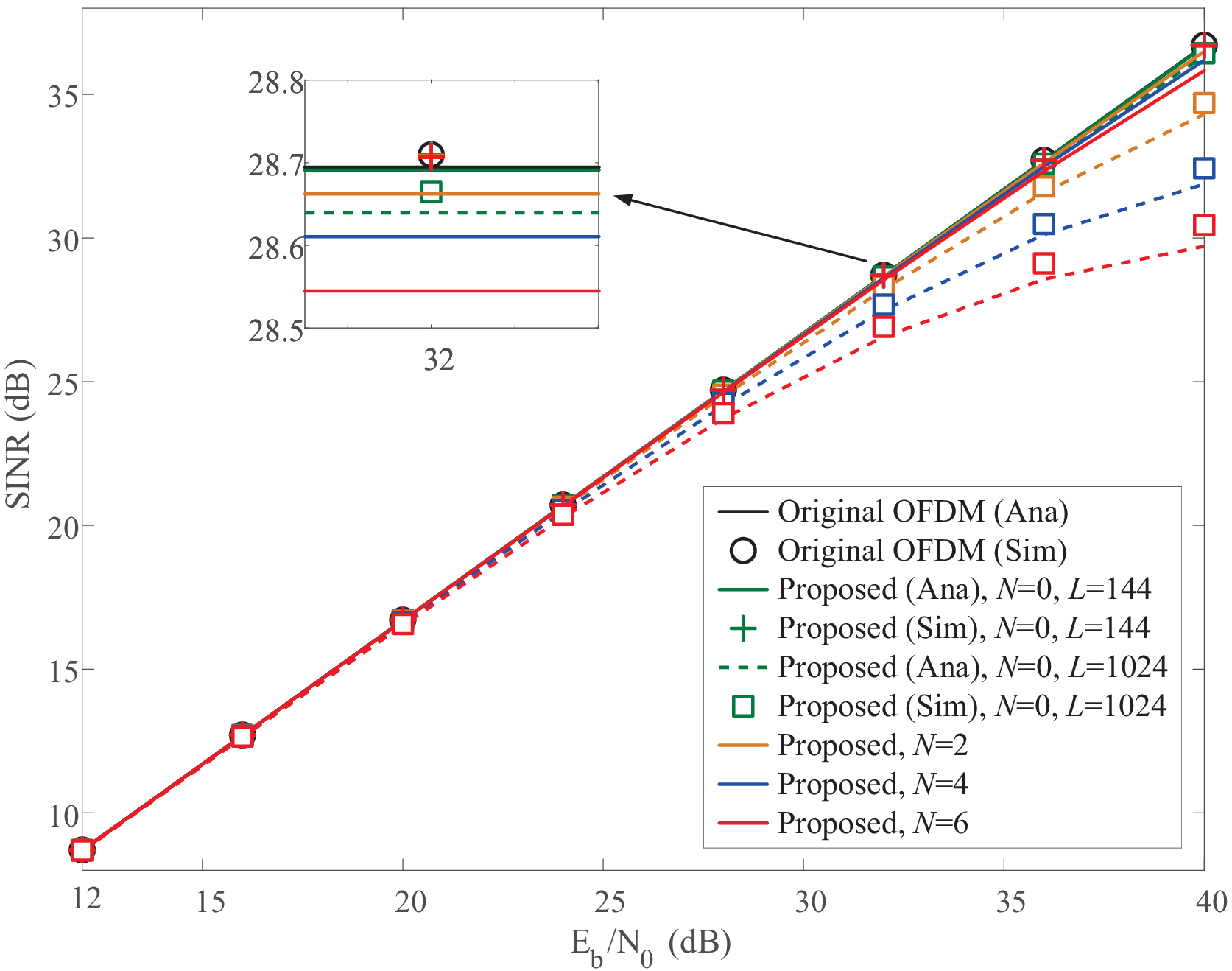}
\caption{Analytical and simulated SNRs of the proposed scheme with different values of \emph{N} and $L$.}
\label{fig:fig2}
\end{figure}

\subsection{Iterative Signal Recovery Algorithm}

The above SNR analysis reveals that the smooth signal in the proposed scheme incurs an SNR loss when the maximum derivative order $N$ is large and the length of $w_i(l)$ significantly exceeds that of the CP. 
To mitigate this interference, we propose the following signal recovery algorithm.

First, substituting \eqref{Eqn1}, \eqref{Eqn:Eqn19}, \eqref{Eqn24}, \eqref{Eqn33}, and \eqref{Eqn34} into \eqref{Eqn31} yields
\begin{align}\label{Eqn42}
   \mathbf{y}_i
   & =    \frac{1}{L_{\rm s}} \mathbf{H}_{i} \mathbf{F}^{\rm H}_1  \mathbf{d}_i - \mathbf{H}_{i} \mathbf{A} \mathbf{U} \mathbf{d}_i + \mathbf{H}_{i} \mathbf{A} \mathbf{d}_{i-1}  + \mathbf{z}_i,
\end{align}
where 
\begin{equation} \label{Eqn44}
\mathbf{A}=  \begin{bmatrix} \begin{bmatrix} \mathbf{0}_{(L-L_{\rm cp}+\tau) \times (L_{\rm cp}-\tau)} & \mathbf{I}_{L-L_{\rm cp}+\tau} \end{bmatrix}  \\ \mathbf{0}_{(L_{\rm s}+L_{\rm cp}-L) \times L} \end{bmatrix}   \left[\mathbf{Q}_f \; \mathbf{Q}_{\tilde{f}} \right] \mathbf{P} \mathbf{P}_1,
\end{equation}
and $\mathbf{F}_1$ is a $K \times (L_{\rm s}+\tau)$ DFT matrix whose $( k,l+\tau+1)$th element is 
$\left\{\mathbf{F}_1 \right\}_{ k,(l+\tau+1)} =  e^{-j\frac{2\pi }{L_{\rm s}}l r_k}$ for $l=-\tau, -\tau+1, \ldots, L_{\rm s}-1$ and $k=1, 2, \ldots, K$.

As shown in Appendix \ref{app:app2}, Eq. \eqref{Eqn44} can be rewritten as
\begin{equation} \label{Eqn40}
  \mathbf{y}_i
   =    \frac{1}{L_{\rm s}} \mathbf{F}^{\rm H} \mathbf{\Lambda}_i  \mathbf{d}_i - \mathbf{H}_{i} \mathbf{A} \mathbf{U} \mathbf{d}_i + \mathbf{H}_{i} \mathbf{A} \mathbf{d}_{i-1}  + \mathbf{z}_i,\end{equation}
where the $K \times K$ diagonal matrix $\mathbf{\Lambda}_{i}$ is given by
\begin{align}  \label{Eqn41}
\mathbf{\Lambda}_{i}  = 
 {\rm diag}\left\{ \epsilon^{-\tau}_{1} g_i(\epsilon_{1}), \epsilon^{-\tau}_{2} g_i(\epsilon_{2}), \ldots, \epsilon^{-\tau}_{K} g_i(\epsilon_{K}) \right\},
\end{align}
with $\epsilon_{k} = e^{j \frac{2\pi }{L_{\rm s}} r_k}$ and $g_i(X)=\sum^{\tau}_{n=0} h_{i,\tau-n}  X^n$.
Note that $\epsilon^{-\tau}_{k} g_i(\epsilon_{k}) =  \sum^{\tau}_{n=0}  h_{i,\tau-n}  \epsilon^{n-\tau}_{k} =  \sum^{\tau}_{n=0}  h_{i,n}  \epsilon^{-n}_{k}  $ represents the frequency-domain channel response on subcarrier $k$. 

We assume perfect channel estimation, i.e., $\mathbf{\Lambda}_{i}$ and $\mathbf{H}_{i}$ are perfectly known as $\tilde{\mathbf{\Lambda}}_{i}$ and $\tilde{\mathbf{H}}_{i}$.
In the $j$th iteration, the smooth signal is removed from the received signal, yielding
\begin{align}\label{Eqn45}
   \bar{\mathbf{y}}_{i,(j)}
  =\mathbf{y}_i  + \tilde{\mathbf{H}}_{i} \mathbf{A} \mathbf{U} \tilde{\mathbf{d}}_{i,(j-1)} - \tilde{\mathbf{H}}_{i} \mathbf{A} \tilde{\mathbf{d}}_{i-1,(j-1)},
\end{align}
where $j=1,2,\ldots,J$ is the iteration index, $J$ denotes the maximum number of iterations, and $\tilde{\mathbf{d}}_{i-1,(j-1)}$ and $\tilde{\mathbf{d}}_{i,(j)}$ represent the symbol-decision vectors for the $(i-1)$th and $i$th OFDM symbols, respectively.

Due to $\mathbf{F} \mathbf{F}^{\rm H} = L_{\rm s} \mathbf{I}_K$, after applying DFT and zero-forcing (ZF) channel equalization, the decision vector becomes
\begin{align}\label{Eqn46}
    \hat{\mathbf{d}}_{i,(j)}
= \tilde{\mathbf{\Lambda}}^{-1}_{i}  \mathbf{F} \bar{\mathbf{y}}_{i,(j)} .
\end{align}
The per-subcarrier symbol decision is expressed as
\begin{equation}\label{Eqn47}
\tilde{d}_{i,r_k,(j)} = \arg \min\limits_{d \in \mathcal{C}} \left\{\left| \hat{d}_{i,r_k,(j)} - d \right|^2\right\},
\end{equation}
where $\mathcal{C}$ is the constellation set, and we initialize $\tilde{d}_{0,r_k,(0)} = 0$ and $\tilde{d}_{1,r_k,(0)} = 0$.

\section{Simulation Results}

We show our performance results employing the simulation parameters listed in Table \ref{tab:tab1}, where since the Blackman window and its first derivative are continuous, we have $N_1=1$.
For comparison, the performance of original OFDM, conventional NC-OFDM \cite{Ref1}, prefix precoding \cite{Ref8}, and prefix/suffix precoding \cite{Ref9} is also considered, where the lengths of the prefix and suffix are set to 72 in the prefix/suffix precoding \cite{Ref9}.
To demonstrate the sidelobe suppression performance, the power spectrum density (PSD) is evaluated by Welch's averaged periodogram method with a 2048-sample Hanning window and 512-sample overlap after observing $10^5$ symbols.

\begin{table} [h]
  \centering
  \caption{Simulation Parameters }
  \label{tab:tab1}
  \begin{tabular}{|c|c|}
\hline
\textbf{Parameters} & \textbf{Values} \\
 \hline
 Constellation modulation  & 16QAM \\
  Number of subcarriers (\emph{K}) & 256    \\
  Subcarrier index set ($\mathcal{R}$) & $\left\{-128,-127,\ldots,127\right\}$    \\
 OFDM symbol duration ($T_{\rm s}$)  & 1/15 ms \\
 Sampling interval ($T_{\rm samp}$) & $T_{\rm s}/2048$    \\
  Frequency interval ($\Delta f$) & 15 KHz \\
  Carrier frequency & 2 GHz \\
  Maximum Doppler shift ($f_D$) & 111.11 Hz \\
  CP length in OFDM ($L_{\rm cp}$) & 144 $T_{\rm samp}$  \\
 Window function ($s(l)$) & Blackman   \\
\hline
\end{tabular}
\end{table}

\subsection{PSD}

\begin{table*} [th]
  \centering
  \caption{Numerical results of ACLR Corresponding to Fig. \ref{fig:fig2}}
  \label{tab:tab2}
  \begin{tabular}{|c|c|c|c|c|}
\hline
\multicolumn{3}{|c|} { \textbf{Scheme}}  & \textbf{ACLR1 (dB)} & \textbf{ACLR2 (dB)}  \\
\hline
\multicolumn{3}{|c|} { \textbf{OFDM}} & 33.76 & 42.43 \\%
\hline
\multirow{4}{*} { \textbf{NC-OFDM \cite{Ref1} }, \textbf{ Precoding \cite{Ref8}}, \textbf{ Precoding \cite{Ref9}} }
  & \multicolumn{2}{c|} { \textbf{\emph{N}=0}} & 39.82, 39.80, 39.68 & 58.57, 58.55, 58.19 \\
\cline{2-5}
  & \multicolumn{2}{c|} { \textbf{\emph{N}=2}} & 50.86, 50.71, 50.11 & 95.10, 95.02, 94.77 \\
\cline{2-5}
  & \multicolumn{2}{c|} { \textbf{\emph{N}=4}} & 60.35, 59.86, 57.88 & 129.23, 128.99, 127.17 \\
\cline{2-5}
  & \multicolumn{2}{c|} { \textbf{\emph{N}=6}} & 69.56, 68.37, 65.15 & 151.53, 151.42, 150.61 \\
\hline
\multirow{4}{*} {\textbf{Proposed}}
& \multirow{4}{60pt} {\textbf{\emph{L}=72, 144, 1024}}
& {\textbf{\emph{N}=0}} & 39.78, 39.53, 39.93 & 58.56, 58.08, 58.33 \\
\cline{3-5}
&& {\textbf{\emph{N}=2}} & 49.08, 50.66, 50.73 & 95.03, 94.97, 94.88 \\
\cline{3-5}
&& {\textbf{\emph{N}=4}} & 51.49, 60.14, 60.36 & 128.48, 129.37, 129.04 \\
\cline{3-5}
&& {\textbf{\emph{N}=6}} & 53.34, 66.61, 69.38 & 150.98, 151.16, 151.25 \\
 \hline
\end{tabular}
\end{table*}

\begin{figure}[h]
\centering
\includegraphics[width=3.5in]{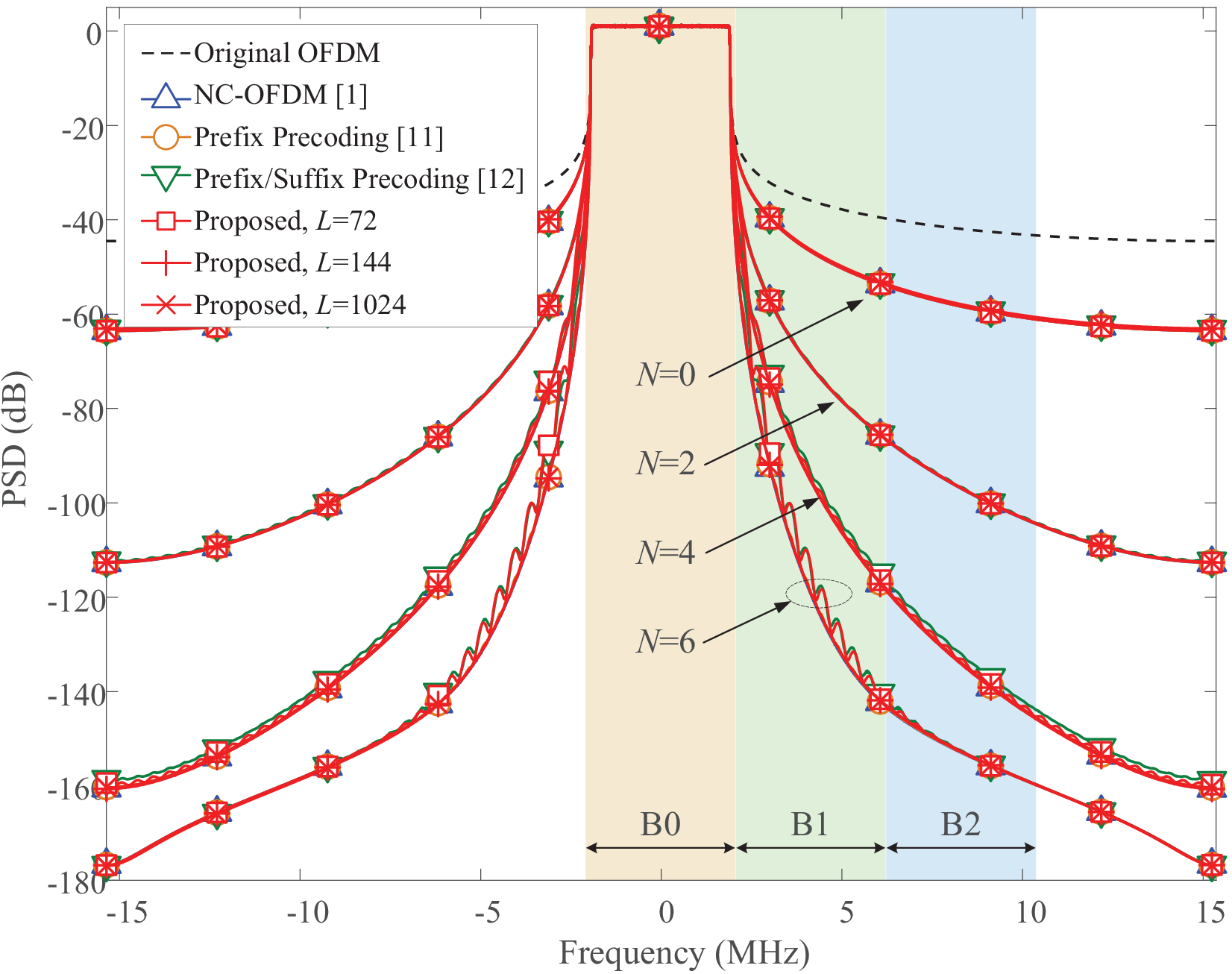}
\caption{PSDs of baseband-equivalent signals for OFDM, NC-OFDM, prefix precoding, prefix/suffix precoding, and the proposed scheme with different $N$ and $L$.}
\label{fig:fig3}
\end{figure}

Fig. \ref{fig:fig3} depicts the sidelobe suppression performance of original OFDM, conventional NC-OFDM, the prefix precoding, the prefix/suffix precoding, and the proposed scheme for \emph{N}=0,2,4,6 and $L$=72,144,1024.
As can be observed, as \emph{N} increases, all \emph{N}-continuous schemes achieve a substantial reduction in out-of-band emission as opposed to OFDM, where NC-OFDM exhibits the best performance.
Moreover, by increasing $L$, the proposed scheme approaches the performance of NC-OFDM.
To quantify this, Table \ref{tab:tab2} lists the adjacent channel leakage power ratio (ACLR) for all the curves in Fig. \ref{fig:fig3}
The ACLR is defined as the ratio of the average power in data band ${\rm B0}$ to the average power outside of band ${\rm B0}$ \cite{Ref12}, where ACLR1 and ACLR2 are associated with band ${\rm B1}$ close to ${\rm B0}$ and band ${\rm B2}$ far from ${\rm B0}$, respectively.
According to the 3GPP standard \cite{Ref19}, the data bandwidth occupies 90\% of the total bandwidth, while the guard bands on both sides account for the remaining 10\%. 
Accordingly, in our simulations, we allocate 14 subcarriers as guard bands on each side of the data band, which together form the frequency set ${\rm B0}$. 
The bands ${\rm B1}$ and ${\rm B2}$ have the same bandwidth as ${\rm B0}$, i.e., ${\rm |B0| = |B1|= |B2|}$=4.26 MHz.
As shown in Table \ref{tab:tab2}, the ACLR of the proposed scheme closely approaches that of NC-OFDM even for small $L$, e.g., $L$=72, especially for ACLR2.
Even in the band ${\rm B1}$, 
For large $N$, such as $N$=4 and 6, the proposed scheme with $L=144$ still attains an ACLR close to that of NC-OFDM in band ${\rm B1}$, confirming that the \emph{N}-continuity is well preserved even with a short-length smooth signal.

\begin{figure}[h]
\centering
\includegraphics[width=3.5in]{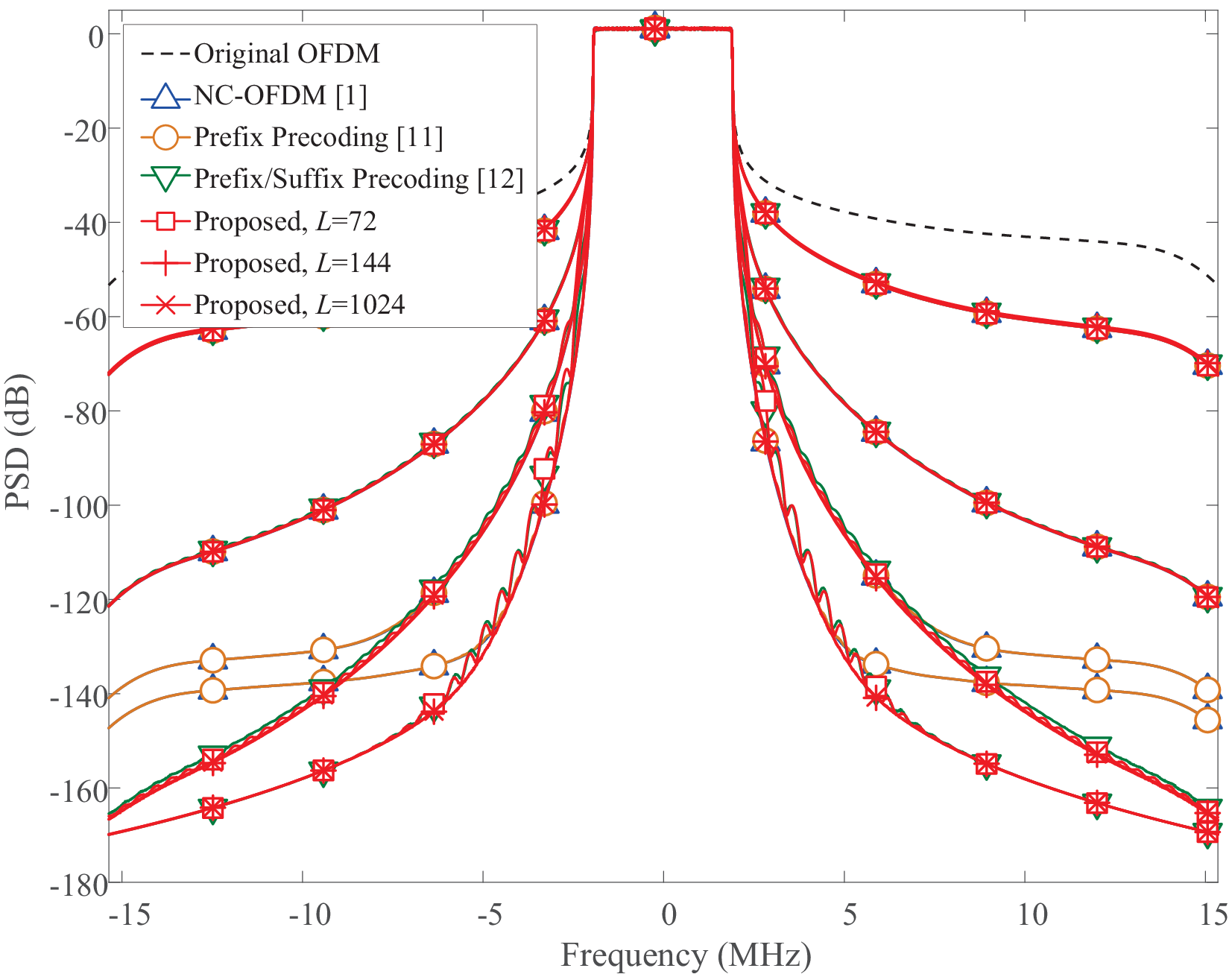}
\caption{PSDs of upsampled and filtered signals for OFDM, NC-OFDM, prefix precoding, prefix/suffix precoding, and the proposed scheme with different $N$ and $L$.}
\label{fig:fig4}
\end{figure}

To further assess the sidelobe suppression in practice, the baseband-equivalent signals corresponding to Fig. \ref{fig:fig3} are upsampled and digitally filtered, and the resulting PSDs are plotted in Fig. \ref{fig:fig4}.
For upsampling, the sampling frequency is set to 122.88 MHz, and a square-root raised cosine pulse with a roll-off factor of 0.25 is interpolated spanning 12 symbols with 4 samples per symbol period.
Then, a low-pass FIR filter of order 80 with a normalized cutoff frequency of 0.25 is applied to suppress imaged spectral components. 
As observed in Fig. \ref{fig:fig4}, the proposed scheme retains a notable reduction in out-of-band emission.
It should be noted that, compared to an ideal Nyquist pulse, a finite-duration non-ideal shaping pulse introduces additional errors, which slightly degrade the smoothness of the \emph{N}-continuous signals. 
Specifically, owing to the edge roll-off to zero, the prefix/suffix precoding and the proposed scheme exhibit only minor spectral leakage regrowth in distant bands for large $N$.
On the contrary, NC-OFDM and the prefix precoding, which lack such roll-off, suffer from more severe out-of-band emission regrowth caused by larger-scale signal discontinuities.

\subsection{BER}

\begin{figure}[h]
\centering
\includegraphics[width=3.5in]{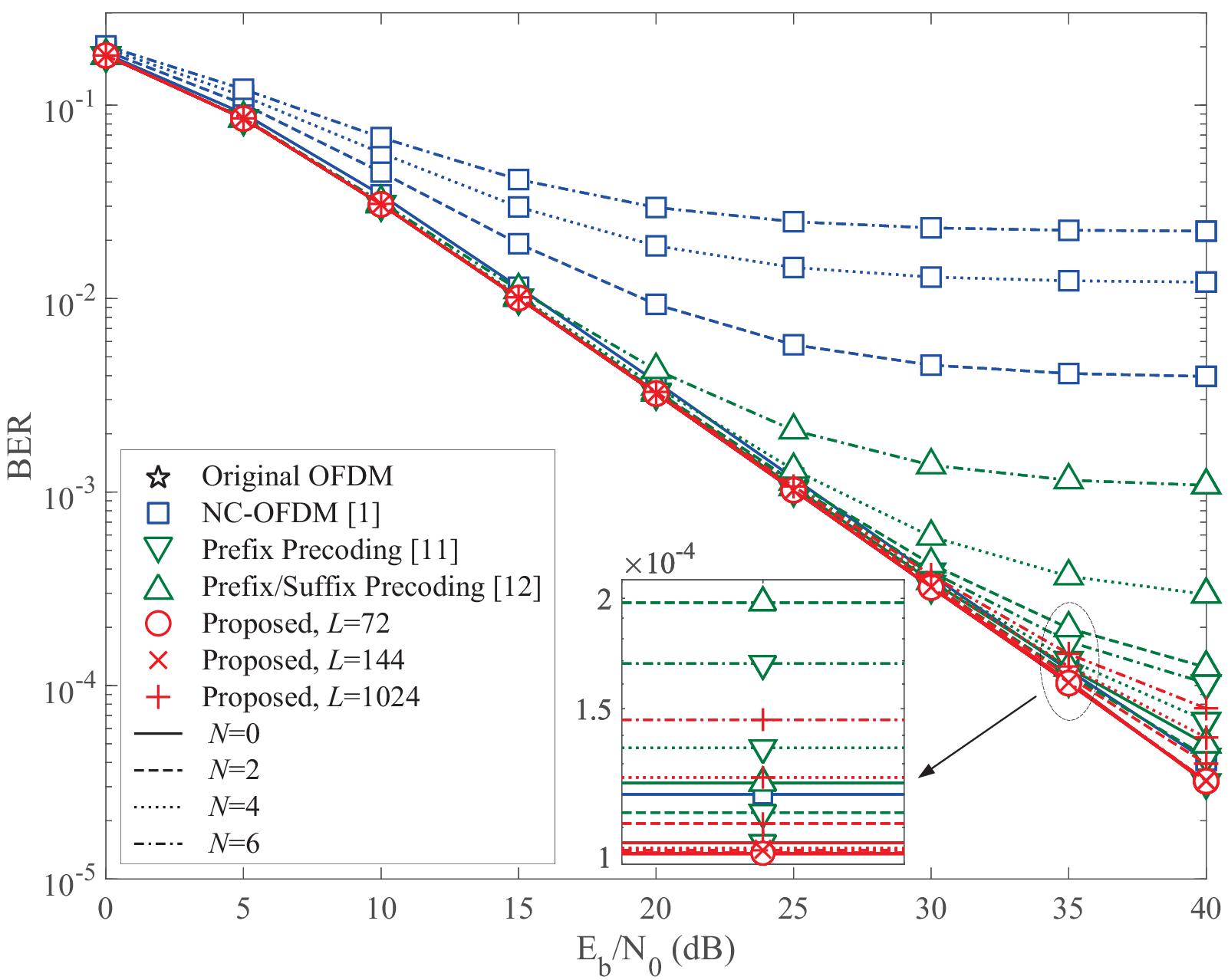}
\caption{BERs of OFDM, NC-OFDM, prefix precoding, prefix/suffix precoding, and the proposed scheme with different \emph{N} and $L$.}
\label{fig:fig5}
\end{figure}

Fig. \ref{fig:fig5} plots the BERs of the proposed scheme and its counterparts over a multipath fading channel, where the channel tap delays and relative powers are set according to the typical urban profile \cite{Ref13} as  [0, 30, 150, 310, 370, 710, 1090, 1730, 2510] ns  [0, -1.5, -1.4, -3.6, -0.6, -9.1, -7, -12, -16.9] dB, respectively. 
The proposed scheme achieves a BER performance close to that of conventional OFDM, while the NC-OFDM receiver exhibits the worst performance, particularly for big \emph{N}, such as \emph{N}=6.
In the low ${\rm E}_{\rm b}/{\rm N}_0$ region, the precoding schemes in \cite{Ref8, Ref9} yield an error performance identical to that of OFDM.
On the contrary, at higher ${\rm E}_{\rm b}/{\rm N}_0$, the interference introduced by these precoders \cite{Ref8,Ref9} has the same duration as the guard interval.
Consequently, for big \emph{N}, e.g., \emph{N}=6 at ${\rm E}_{\rm b}/{\rm N}_0$=35 dB, their BER performance is noticeably degraded by the channel-delayed self-interference.
It can be also inferred that the proposed scheme's performance is not affected by the smooth signal, owing to the fact that the proposed scheme is capable of adaptively reducing the length of the smooth signal so that the channel-delayed smooth signal cannot interfere with the useful signal.

\begin{figure}[thpb]
	\centering
	\includegraphics[width=3.5in]{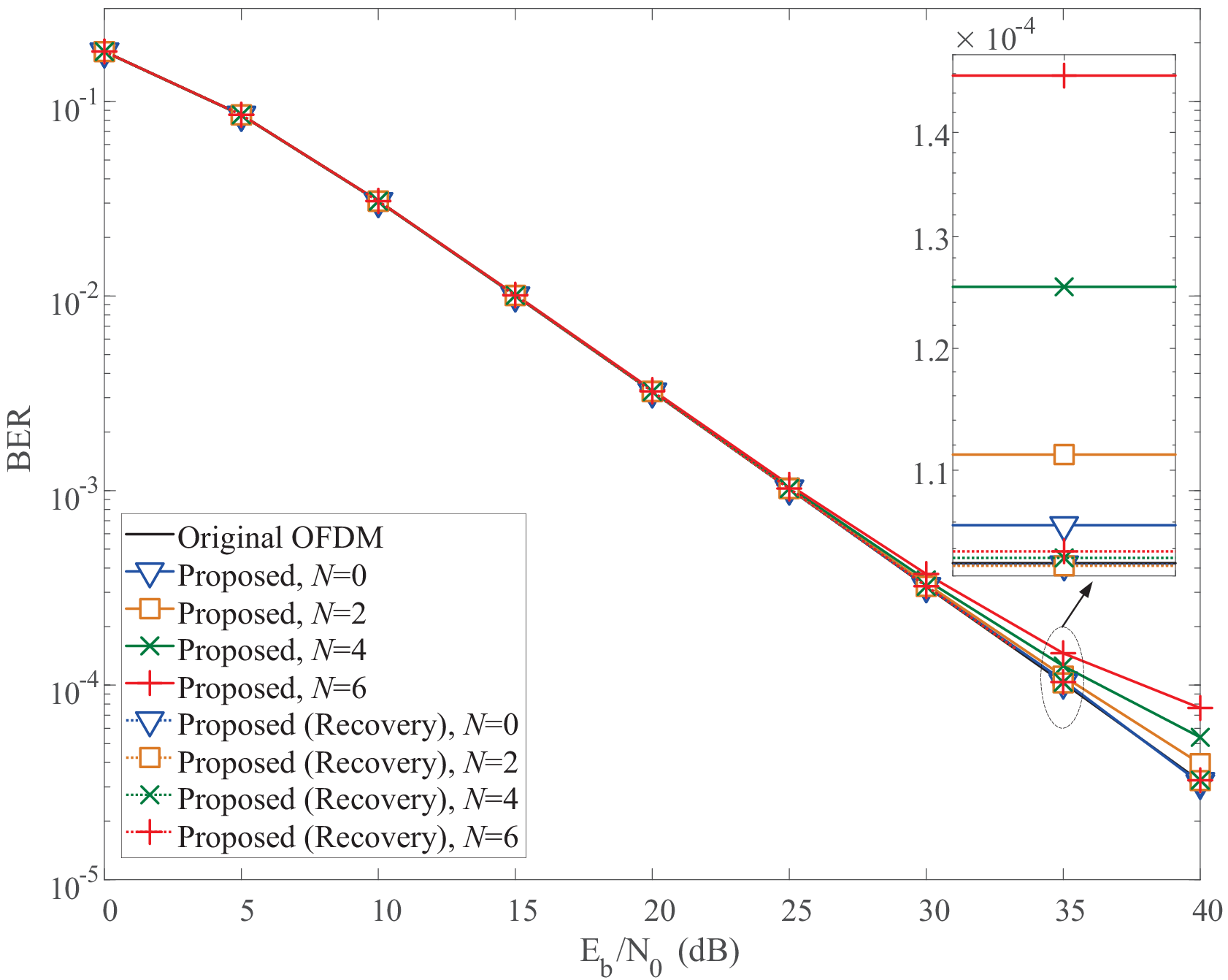}
	\caption{BERs of the proposed scheme with the signal recovery.}
	\label{fig:fig6}
\end{figure}

Fig. \ref{fig:fig6} validates the effectiveness of the proposed signal recovery algorithm presented in Section IV-B when applied to the proposed scheme with a long smooth signal, where the number of iteration is set as $J$=2.
Consistent with the SNR and BER analyses in Fig. \ref{fig:fig2} and Fig. \ref{fig:fig5}, we set $L$=1024 as a representative long length of the smooth signal.
It can be seen that the signal recovery algorithm attains a BER closely approaching that of original OFDM.
Due to the high complexity of the signal recovery algorithm, adaptively adjusting the length of the smooth signal constitutes a more practical choice when balancing sidelobe suppression against BER performance.

\subsection{Complexity}

\begin{figure}[thpb]
	\centering
	\includegraphics[width=3.5in]{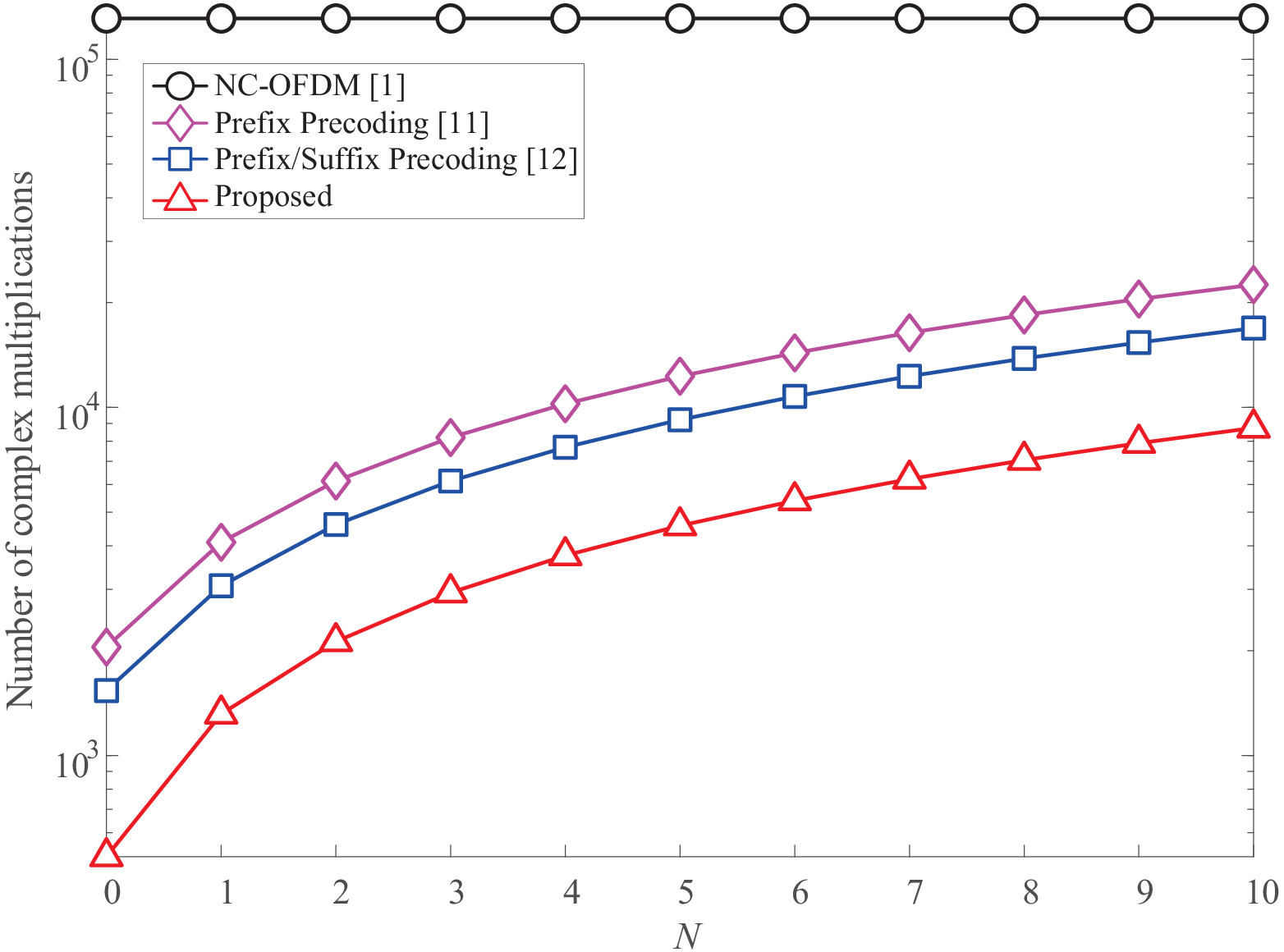}
	\caption{Numerical comparison of number of complex multiplications at the transmitter in NC-OFDM, prefix precoding, prefix/suffix precoding, and the proposed scheme with $L$=144.}
	\label{fig:fig7}
\end{figure}

Table \ref{tab:tab3} compares the additional complexity of the proposed scheme and its counterparts relative to conventional OFDM, quantified by the number of complex multiplications. 
Based on \eqref{Eqn:Eqn19}, \eqref{Eqn20}, and \eqref{Eqn27}, the term ${\mathbf{P}}_1\mathbf{d}_{i-1}-\mathbf{P}_2\mathbf{d}_i$ requires $2(N+1)K$ complex multiplications, $\mathbf{P} \Delta \mathbf{x}$ requires $(N+1)(N+\tilde{N}+2)$ complex multiplications, and $\mathbf{Q}_f \mathbf{b}_i + \mathbf{Q}_{\tilde{f}} \tilde{\mathbf{b}}_i$ requires $L(N+\tilde{N}+2)$ complex multiplications. 
As shown in Figs. \ref{fig:fig3}--\ref{fig:fig6} and Table \ref{tab:tab2}, the proposed scheme achieves favorable sidelobe suppression and BER performance with $L$ much smaller than $K=256$ (e.g., $L=144$). 
The use of a Blackman window for $s_1(l)$ yields $N_1=1$. 
In this case, Fig. \ref{fig:fig7} shows that the proposed scheme has the lowest complexity among all $N$-continuous methods. 
Specifically, for $N=4$, the reduced complexity ratios of the proposed scheme to NC-OFDM, prefix precoding \cite{Ref8}, and prefix/suffix precoding \cite{Ref9} are 97.14\%, 63.36\%, and 51.15\%, respectively.

\begin{table} [!htpb]
 \centering
  \caption{Additional Complexity of Various \emph{N}-Continuous Schemes Compared to Original OFDM}
  \label{tab:tab3}
  \begin{tabular}{|m{53pt}<{\centering}|m{155pt}<{\centering}|}
\hline
\textbf{Scheme} & \textbf{Number of complex multiplications}  \\
\hline
\textbf{NC-OFDM \cite{Ref1} }  & $2K^2$ \\
\hline
\textbf{Precoding \cite{Ref8} } & $8(N+1)K$  \\
\hline
\textbf{Precoding \cite{Ref9} } & $6(N+1)K$  \\
\hline
\textbf{Proposed}  & $2(N+1)K+(2N+1-N_1)(L+N+1)$ \\
\hline
\end{tabular}
\end{table}

\section{Conclusion}

In this paper, we proposed a class of novel \emph{N}-continuous OFDM schemes designed for interference cancellation, through enhancing the continuities of the basis signals and their higher-order derivatives.
Our simulation results revealed that the proposed scheme outperforms its counterparts in the trade-off among sidelobe suppression, computational complexity and BER performance.
Specifically, with a short-length smooth signal, the proposed scheme maintains a BER close to that of conventional OFDM while imposing a lower complexity than existing low-interference NC-OFDM precoders and attaining a sidelobe suppression performance comparable to that of conventional NC-OFDM.


\appendices

\renewcommand{\theequation}{\thesection.\arabic{equation}}

\section{Proof of \eqref{Eqn39}}
\label{app:app1}
\setcounter{equation}{0}

In \eqref{Eqn39}, $  {\rm Tr} \left\{ E\left\{ \mathbf{H}_{d_i} \mathbf{F}^{\rm H} \mathbf{d}_i \mathbf{d}^{\rm H}_i \mathbf{F} \mathbf{H}^{\rm H}_{d_i} \right\} \right\}$ can be expressed as
\begin{align} \label{AppB:Eqn1}
{\rm Tr} &\left\{ E\left\{ \mathbf{H}_{d_i} \mathbf{F}^{\rm H} \mathbf{d}_i \mathbf{d}^{\rm H}_i \mathbf{F} \mathbf{H}^{\rm H}_{d_i} \right\} \right\}  \nonumber \\
& = E \left\{ {\rm Tr} \left\{  \mathbf{F}^{\rm H} \mathbf{d}_i \mathbf{d}^{\rm H}_i \mathbf{F} \mathbf{H}^{\rm H}_{d_i} \mathbf{H}_{d_i}\right\} \right\}  \nonumber \\
& = {\rm Tr} \left\{    \mathbf{F}^{\rm H} E \left\{\mathbf{d}_i \mathbf{d}^{\rm H}_i\right\}  \mathbf{F} E\left\{\mathbf{H}^{\rm H}_{d_i} \mathbf{H}_{d_i}\right\} \right\}  \nonumber \\
& = E\left\{{\rm Tr} \left\{    \mathbf{F}^{\rm H}  \mathbf{F} \mathbf{H}^{\rm H}_{d_i} \mathbf{H}_{d_i}\right\} \right\}  .
\end{align}

It is assumed that $\bar{\mathbf{F}}$ is an $L_{\rm s} \times L_{\rm s}$ DFT matrix with its element of $\left\{\bar{\mathbf{F}}\right\}_{l+1,\bar{l}+1}  = e^{-j 2 \pi \frac{l}{L_{\rm s}}\bar{l}}$ for $l,\bar{l} = 0,1, \ldots,L_{\rm s}-1$. 
According to the property of the circulant matrix, $\mathbf{H}_{d_i}$ can be expressed as $\mathbf{H}_{d_i} = \frac{1}{L_{\rm s}} \bar{\mathbf{F}}^{\rm H} \mathbf{\Sigma}_i \bar{\mathbf{F}}$, where $\mathbf{\Sigma}_i = {\rm diag} \left\{ \lambda_{i,1}, \lambda_{i,2}, \ldots,\lambda_{i,L_{\rm s}} \right\}$ is an $L_{\rm s} \times L_{\rm s}$ diagonal matrix, and the $(l+1)$th eigenvalue $\lambda_{i,l+1}$ is expressed as $\lambda_{i,l+1}= \sum\limits^{\tau}_{u=0} h_{i,u} e^{-j 2 \pi \frac{u}{L_{\rm s}} l}$.
In fact, $\lambda_{i,l+1}$ is the frequency-domain channel response.
Thus, we have 
\begin{equation} \label{AppB:Eqn2}
\mathbf{H}^{\rm H}_{d_i} \mathbf{H}_{d_i} = \frac{1}{L_{\rm s}} \bar{\mathbf{F}}^{\rm H} \mathbf{\Sigma}^{\rm H}_i  \mathbf{\Sigma}_i \bar{\mathbf{F}} .
\end{equation}
Meanwhile, $\mathbf{F}^{\rm H}  \mathbf{F}$ can be expressed as 
\begin{align} \label{AppB:Eqn3}
\mathbf{F}^{\rm H}  \mathbf{F} = \bar{\mathbf{F}}^{\rm H} \mathbf{\Sigma} \bar{\mathbf{F}},
\end{align}
where $\mathbf{\Sigma}$ is a diagonal matrix with $K$ entries equal to 1 on its diagonal, located at the positions corresponding to the subcarriers in the set $\mathcal{R}$.

Based on \eqref{AppB:Eqn2} and \eqref{AppB:Eqn3}, we can arrive at 
\begin{align} \label{AppB:Eqn4}
{\rm Tr} \left\{    \mathbf{F}^{\rm H}  \mathbf{F} \mathbf{H}^{\rm H}_{d_i} \mathbf{H}_{d_i}\right\} 
& =  \frac{1}{L_{\rm s}} {\rm Tr} \left\{  \bar{\mathbf{F}}^{\rm H} \mathbf{\Sigma} \bar{\mathbf{F}} \bar{\mathbf{F}}^{\rm H} \mathbf{\Sigma}^{\rm H}_i  \mathbf{\Sigma}_i \bar{\mathbf{F}} \right\}  \nonumber \\
& =  \frac{1}{L_{\rm s}} {\rm Tr} \left\{   \mathbf{\Sigma} \bar{\mathbf{F}} \bar{\mathbf{F}}^{\rm H} \mathbf{\Sigma}^{\rm H}_i  \mathbf{\Sigma}_i \bar{\mathbf{F}} \bar{\mathbf{F}}^{\rm H} \right\} \nonumber \\
&= L_{\rm s} {\rm Tr} \left\{   \mathbf{\Sigma} \mathbf{\Sigma}^{\rm H}_i  \mathbf{\Sigma}_i \right\} \nonumber \\
& = L_{\rm s} \sum\limits^{K}_{k=1} |\lambda_{i,r_k}|^2.
\end{align}

Therefore, Eq. \eqref{AppB:Eqn1} can be expressed as 
\begin{align} \label{AppB:Eqn5}
{\rm Tr} \left\{ E\left\{ \mathbf{H}_{d_i} \mathbf{F}^{\rm H} \mathbf{d}_i \mathbf{d}^{\rm H}_i \mathbf{F} \mathbf{H}^{\rm H}_{d_i} \right\} \right\}  
 =  L_{\rm s} \sum\limits^{K}_{k=1} E\left\{ |\lambda_{i,r_k}|^2 \right\} .
\end{align}

Then, according to \eqref{Eqn:Eqn19}, \eqref{Eqn22}, and \eqref{Eqn24}, ${\rm Tr} \left\{E\left\{ \mathbf{H}_{w_i} \mathbf{w}_i \mathbf{w}^{\rm H}_i \mathbf{H}^{\rm H}_{w_i} \right\}\right\} $ in \eqref{Eqn39} can be expressed as 
\begin{align}\label{AppB:Eqn6}
{\rm Tr} & \left\{E\left\{ \mathbf{H}_{w_i} \mathbf{w}_i \mathbf{w}^{\rm H}_i \mathbf{H}^{\rm H}_{w_i} \right\}\right\}   \nonumber \\
&= {\rm Tr} \left\{E\left\{ \mathbf{H}^{\rm H}_{w_i}  \mathbf{H}_{w_i} \right\} E\left\{ \mathbf{w}_i \mathbf{w}^{\rm H}_i \right\}\right\}  \nonumber \\
&= {\rm Tr} \bigg\{ 
E\left\{ \mathbf{H}^{\rm H}_{w_i}  \mathbf{H}_{w_i} \right\}    
 \begin{bmatrix}  \mathbf{Q}_{{f}}  &\!\!\! \mathbf{Q}_{\tilde{f}}  \end{bmatrix} \mathbf{P}  
 \big(  \mathbf{P}_1 E\left\{\mathbf{d}_{i-1} \mathbf{d}^{\rm H}_{i-1} \right\}  \mathbf{P}^{\rm H}_1  \nonumber \\
 & \quad + \mathbf{P}_2 E\left\{\mathbf{d}_{i} \mathbf{d}^{\rm H}_{i} \right\}  \mathbf{P}^{\rm H}_2 \big)  
 \mathbf{P}^{\rm H}   \begin{bmatrix}  \mathbf{Q}^{\rm H}_{{f}}  \\  \mathbf{Q}^{\rm H}_{\tilde{f}}  \end{bmatrix} 
 \bigg\}  \nonumber \\
 &= 2 {\rm Tr} \left\{ 
E\left\{ \mathbf{H}^{\rm H}_{w_i}  \mathbf{H}_{w_i} \right\}    
 \begin{bmatrix}  \mathbf{Q}_{{f}}  &\!\!\! \mathbf{Q}_{\tilde{f}}  \end{bmatrix} \mathbf{P}  
  \mathbf{P}_1   \mathbf{P}^{\rm H}_1  \mathbf{P}^{\rm H}   \begin{bmatrix}  \mathbf{Q}^{\rm H}_{{f}}  \\  \mathbf{Q}^{\rm H}_{\tilde{f}}  \end{bmatrix} 
 \right\} .
\end{align}

Finally, substituting \eqref{AppB:Eqn5} and \eqref{AppB:Eqn6} into \eqref{Eqn38} yields \eqref{Eqn39}.

\section{Proof of $\mathbf{H}_i \mathbf{F}^{\rm H}_1 =  \mathbf{F}^{\rm H} \mathbf{\Lambda}_i$}
\label{app:app2}   
\setcounter{equation}{0}

Upon generalizing the properties of the circulant matrix in \cite{Ref14,Ref15},  we obtain the following results of $\mathbf{H}_i$ in \eqref{Eqn30}.

\begin{lemma} \label{Lem:lem1}
For any $n$th root $\epsilon^{n}_{k} = e^{j \frac{2\pi }{L_{\rm s}} r_k n}$ of unity and any polynomial $g_i(X)=\sum\limits^{\tau}_{n=0} g_{i,n} X^n$ over $\mathbb{F}$ with degree ${\rm deg} \{g_i(X)\} \leqslant \tau$,
\begin{equation} \label{AppA:Eqn1}
\mathbf{H}_i \mathbf{e}_{k}= \epsilon^{-\tau}_{k} g_i(\epsilon_{k}) \bar{\mathbf{e}}_{k},  
\end{equation}
where $\mathbb{F}$ is a field whose characteristic ${\rm char} \{ \mathbb{F} \}$ is zero or coprime to $L_{\rm s}$,  $r_k\in \mathcal{R}$ with $k=1,2,\ldots,K$,
\begin{align} \label{AppA:Eqn2}
\mathbf{e}_{k}
= \begin{bmatrix} \epsilon^{-\tau}_{k} & \epsilon^{-\tau+1}_{k} & \ldots & \epsilon^{0}_{k} & \ldots & \epsilon^{L_{\rm s}-1}_{k}  \end{bmatrix}^{\rm T},  
\end{align}
\begin{equation}\label{AppA:Eqn3}
\bar{\mathbf{e}}_{k}
= \begin{bmatrix} \epsilon^{0}_{k} & \epsilon^{1}_{k} & \ldots & \epsilon^{L_{\rm s}-1}_{k}  \end{bmatrix}^{\rm T}.
\end{equation}

\end{lemma}

\begin{proof}
The first row of $\mathbf{H}_i \mathbf{e}_{k}$ can be expressed as
\begin{align} 
 \begin{bmatrix} h_{i,\tau} & \cdots & h_{i,0} & 0 & \cdots & 0 \end{bmatrix} 
\mathbf{e}_{k}  
& =  \epsilon^{-\tau}_{k} \sum\limits^{\tau}_{n=0}  h_{i,\tau-n}  \epsilon^{n}_{k}  \nonumber \\
& = \epsilon^{-\tau}_{k} g_i(\epsilon_{k}), \nonumber 
\end{align}
where $g_{i,n}=h_{i,\tau-n} $.
Then, the second row of $\mathbf{H}_i \mathbf{e}_{k}$ can be expressed as
\begin{align} 
 \begin{bmatrix} 0 \!&\! h_{i,\tau} \!&\!\! \cdots \!\!&\! h_{i,0} \!&\! 0 \!&\!\! \cdots \!\!&\! 0 \end{bmatrix} 
\mathbf{e}_{k}  
& =  \epsilon^{-\tau+1}_{k} \sum\limits^{\tau}_{n=0} h_{i,\tau-n}  \epsilon^{n}_{k}   \nonumber \\
& = \epsilon^{-\tau+1}_{k} g_i(\epsilon_{k}).  \nonumber 
\end{align}
Thus, for $m=1,2,\ldots,L_{\rm s}$, the $m$th row of $\mathbf{H}_i \mathbf{e}_{k}$ is given by
\begin{align} 
 \epsilon^{-\tau+m-1}_{k} \sum\limits^{\tau}_{n=0} h_{i,\tau-n} \epsilon^{n}_{k} = \epsilon^{-\tau+m-1}_{k} g_i(\epsilon_{k}).  \nonumber 
\end{align}

Finally, based on the mathematical induction, we obtain
\begin{equation} \label{AppA:Eqn4}
\mathbf{H}_i \mathbf{e}_{k}= \epsilon^{-\tau}_{k} g_i(\epsilon_{k})  \begin{bmatrix} 1 & \epsilon_{k} & \cdots & \epsilon^{L_{\rm s}-1}_{k} \end{bmatrix}^{\rm T}.
\end{equation}
Thus, Lemma \ref{Lem:lem1} is proved.

\hfill{$\blacksquare$}
\end{proof}

Based on Lemma \ref{Lem:lem1}, we further derive the following Lemma \ref{Lem:lem2}.

\begin{lemma} \label{Lem:lem2}
For any polynomial $g_i(X)=\sum\limits^{\tau}_{n=0} g_{i,n} X^n$ over $\mathbb{F}$ with degree ${\rm deg} \{g_i(X)\} \leqslant \tau$, we have
\begin{equation} \label{AppA:Eqn5}
\mathbf{H}_i \mathbf{F}^{\rm H}_1 =  \mathbf{F}^{\rm H} \mathbf{\Lambda}_i,
\end{equation}
where $\mathbf{F}_1$ is a $K \times (L_{\rm s}+\tau)$ DFT matrix given by
\begin{equation} \label{AppA:Eqn6}
\left\{\mathbf{F}_1 \right\}_{k,(m+\tau+1) } =  e^{-j\frac{2\pi }{L_{\rm s}}m r_k},
\end{equation}
for $m=-\tau, -\tau+1, \ldots, L_{\rm s}-1$, $k=1, 2, \ldots, K$, and $r_k \in \mathcal{R}$.
\end{lemma}

\begin{proof}
According to the block multiplication of block matrices and \eqref{AppA:Eqn1}, we have
\begin{align}
 \mathbf{H}_i \mathbf{F}^{\rm H}_1  
&=  \mathbf{H}_i   \begin{bmatrix}  \mathbf{e}_{1} & \mathbf{e}_{2} & \cdots & \mathbf{e}_{K}  \end{bmatrix}   \nonumber \\
&= \begin{bmatrix} \epsilon^{-\tau}_{1} g_i(\epsilon_{1}) \bar{\mathbf{e}}_{1}  & \epsilon^{-\tau}_{2} g_i(\epsilon_{2}) \bar{\mathbf{e}}_{2}  &  \cdots &  \epsilon^{-\tau}_{K} g_i(\epsilon_{K}) \bar{\mathbf{e}}_{K} \end{bmatrix}  \nonumber \\
&= \begin{bmatrix}    \bar{\mathbf{e}}_{1}  &   \bar{\mathbf{e}}_{2} & \cdots &   \bar{\mathbf{e}}_{K}  \end{bmatrix}  \nonumber \\
&\quad  \cdot {\rm diag}\left\{ \epsilon^{-\tau}_{1} g_i(\epsilon_{1}), \epsilon^{-\tau}_{2} g_i(\epsilon_{2}), \ldots, \epsilon^{-\tau}_{K} g_i(\epsilon_{K}) \right\}  \nonumber \\
&= \mathbf{F}^{\rm H} \mathbf{\Lambda}_i. \nonumber
\end{align}
\hfill{$\blacksquare$}
\end{proof}

Thus, \eqref{AppA:Eqn5} is proved.

\end{document}